\documentclass{article}

\newcommand{\be}{\begin{equation}}
\newcommand{\ee}{\end{equation}}

\newcommand{\bea}{\begin{eqnarray}}
\newcommand{\eea}{\end{eqnarray}}
\newcommand{\ba}{\begin{array}}
\newcommand{\ea}{\end{array}}

\def\a{\alpha}
\def\b{\beta}
\def\g{\gamma}
\def\c{\chi}

\def\e{\epsilon}

\def\f{\phi}

\def\k{\kappa}

\def\m{\mu}

\def\p{\pi}

\def\r{\rho}
\def\s{\sigma}

\def\F{\Phi}

\newcommand{\ns}{\normalsize}

\begin{document}

\title{
\hfill{\ns SUSX-TH/02-022\\}
\hfill{\ns hep-th/0210026\\[2cm]}
Moving Five-Branes and Cosmology~\footnote{Based on a talk given
at the 1st International Conference on String Phenomenology, Oxford,
6-11 Jul 2002.}}

\author{Andr\'e Lukas \\[0.2cm]
{\ns Centre for Theoretical Physics} \\
{\ns University of Sussex} \\ 
{\ns Falmer, Brighton BN1 9QJ, United Kingdom} \\[0.2cm] 
{\ns E-mail: a.lukas@sussex.ac.uk}}

\date{}

\maketitle


\begin{abstract}
We discuss low-energy heterotic M-theory with five-branes in four and
five dimensions and its application to moving brane cosmology.
\end{abstract}


\section{Introduction}

If string/M-theory is the correct fundamental theory of nature
time-evolution of branes is likely to have played an important role in the
history of the early universe. In this talk, I will review several
aspects of this early universe brane dynamics in the context of
strongly-coupled heterotic string-theory. This theory can be
formulated as 11-dimensional supergravity on the orbifold $S^1/Z_2$
coupled to two 10-dimensional $E_8$ super-Yang-Mills multiplets
localised on the two orbifold fixed planes (or
boundaries)~\cite{Horava:1996ma}. Upon compactification on a
Calabi-Yau three-fold one arrives at a five-dimensional $N=1$
supergravity theory on $S^1/Z_2$ where the now four-dimensional
boundaries carry $N=1$ gauge-theories which originate from the
10-dimensional Yang-Mills theories~\cite{Lukas:1998yy,Brandle:2001ts}.
Further, M-theory five-branes can be included in this
compactification. They wrap two-cycles within the internal Calabi-Yau
space and, hence, appear as three-branes from the viewpoint of the
effective five-dimensional theory. It is the dynamics of these M
five-branes/three-branes which we are going to focus on in the following.

It should be stressed that our approach to early universe cosmology has
the virtue of being embedded in a meaningful and successful environment
for particle phenomenology from M-theory. Specifically, one can construct
models within the class of compactifications described above which
lead to phenomenologically promising theories close to the standard
model of particle physics located on one of the
boundaries~\cite{Donagi:1999xe,Donagi:2000zf}.

The plan of this talk is as follows. To set the scene, I will present
the effective five- and four-dimensional theories of heterotic
M-theory including three-branes in their minimal versions. The status
and interpretation of these effective theories as well as their general
implications for three-brane dynamics will be reviewed.
Subsequently, specific classes of solutions, both in four and five
dimensions, describing simple three-brane evolution will be
presented. Finally, we will discuss a new approach to brane-dynamics
where three-branes are modelled as topological defects (kinks) of a
bulk scalar field theory. Within this approach, more complicated
brane evolution can be analysed by studying the dynamics of the
modelling bulk scalar field. Particularly, what amounts to
topology-changing transitions of the original M-theory action can be
explicitly  described by certain transitions within this bulk scalar field
theory. 

\section{Effective actions in five and four dimensions}

The five-dimensional effective action for heterotic M-theory is given
by a specific $D=5$, $N=1$ gauged supergravity in the bulk coupled to
four-dimensional $N=1$ gauge theories on the boundaries and the
three-branes. In its minimal version there are two bulk
supermultiplets, namely the supergravity multiplet containing the
metric, the gravi-photon and the gravitini and the universal
hypermultiplet containing the dilaton $\F$ along with three
pseudo-scalar partners and the associated fermions. On the boundaries
$M_4^i$, where $i=1,2$, one obtains $N=1$ gauge theories with gauge groups
$G_i\subset E_8$ and chiral matter fields transforming under
$G_i$. The details of this boundary field content depend on the
compactification from eleven to five dimensions which is the main
subject of particle physics model building in heterotic M-theory. Each
three-brane $M_4^3$ (and its $Z_2$ mirror $\tilde{M}_4^3$) carries on
its world-volume a universal chiral multiplet containing the field $Y$,
specifying the three-brane position in the transverse space, along
with an axionic partner and $U(1)$ vector multiplets. The number of
these vector multiplets is given by the genus $g$ of the curve wrapped
by the M five-brane.

The effective action for these fields~\cite{Lukas:1998yy,Brandle:2001ts}
can be consistently truncated to an even simpler version containing
the metric $g_{\a\b}$ and the dilaton $\F$ as the only bulk fields and the
embedding coordinates $X^\a = X^\a(\s^\m )$ as the only brane fields.
For a single three-brane, the associated effective action is then given by
\begin{eqnarray}
 S_5 &=& -\frac{1}{2\k_5^2}\left\{\int_{M_5}\sqrt{-g}\left[
         \frac{1}{2}R+\frac{1}{4}\partial_\a\F\partial^\a\F
         +\frac{1}{3}\a^2 e^{-2\F}\right]\right. \nonumber \\
     &&\qquad +\int_{M_4^1}\sqrt{-g}\; 2\a_1 e^{-\F} 
       +\int_{M_4^2}\sqrt{-g}\; 2\a_2 e^{-\F} \nonumber \\
     &&\qquad \left. +\int_{M_4^3\cup\tilde{M}_4^3}\sqrt{-\g}\;
       \a_3 e^{-\F}\right\}\; . \label{S5}
\end{eqnarray}
where $\a_i=\s\b_i$, $\b_i\in {\bf Z}$ for $i=1,2,3$ are the charges
on the boundaries and the three-brane quantised in units of $\s$.
These charges satisfy the cohomology condition
\begin{equation}
 \sum_{i=1}^3\a_i = 0 \label{cohomology}\; .
\end{equation}
The gauge-charge $\a$ is defined as a sum of step-functions
\begin{equation}
 \a = \a_1\theta (M_4^1) + \a_2\theta (M_4^2) + \a_3\left(\theta (M_4^3)+
      \theta (\tilde{M}_4^3)\right) \; . \label{alpha}
\end{equation}
Note that two actions of the above type with two different sets of charges
$\a_i$ correspond to topologically different M-theory compactifications.

The four-dimensional effective action obtained from~(\ref{S5}) by
compactifying on the orbifold is given by
\begin{equation}
 S_4 = 
-\frac{1}{2\k_P^2}\int_{M_4}\sqrt{-g_4}\left[\frac{1}{2}R_4+
 \frac{1}{4}\partial_\m\f\partial^\m\f +\frac{3}{4}\partial_\m\b\partial^\m\b
 +\frac{q_3}{2}e^{\b -\f}\partial_\m z\partial^\m z\right]\;  \label{S4}
\end{equation}
where $q_3=\p\r\a_3$. Here, the field $\f$, as the zero mode of the
five-dimensional dilaton $\F$ measures the Calabi-Yau size, while the
field $\b$ originates from the $(55)$ component of the metric and
measures the orbifold size. Finally, the field $z$ specifies the
three-brane position and is normalised so that $z\in [0,1]$ with
$z=0,1$ corresponding to the two boundaries. This effective action
can be obtained from the K\"ahler potential~\cite{Derendinger:2001gy}
\begin{equation}
 K = -\ln\left( S+\bar{S}-q_3\frac{(Z+\bar{Z})^2}{T+\bar{T}}\right)
     -3\ln\left( T+\bar{T}\right)\; , \label{K}
\end{equation}
by truncating off the axionic fields contained in the chiral
superfields $S\leftrightarrow \f$, $T\leftrightarrow \b$ and
$Z\leftrightarrow z$.  Perturbatively, these moduli superfields are
flat directions but one expects non-perturbative contributions to
their superpotential from a number of
sources~\cite{Moore:2000fs,Lima:2001nh}. Assuming the axions can still
be integrated out, these contributions can be included by adding a
suitable potential $V=V(\f ,\b ,z)$ to the above action. Note that
this potential will be a function of all three fields rather than $z$
alone.  It is quite clear even from the free action~(\ref{S4}) that
three-brane evolution in a static geometry (that is, for constant $\f$
and $\b$) is not possible. For this reason, the above action (possibly
with a potential) represents the minimal system in which heterotic
moving-brane cosmology can be discussed in a meaningful way.

\section{Cosmological solutions with a moving three-brane}

The most general cosmological solutions of~(\ref{S4}) with flat
spatial sections~\cite{Copeland:2001zp} show a number of interesting
generic features.  Asymptotically at early and late time, both in the
negative and positive time branch, the three-brane is practically at
rest while the dilaton $\f$ and the T-modulus $\b$ evolve according to
a standard rolling radii solution. However, the early and late rolling
radii solutions are generally different. Interpolation between these
rolling radii solutions is achieved at intermediate time due to
non-trivial evolution of the three-brane. It turns out that not all
rolling radii solutions can be approached asymptotically but only
those for which the strong-coupling expansion parameter
\begin{equation}
 \e \sim e^{\b -\f}
\end{equation}
diverges at early and late time. The brane-evolution, therefore, drives
the system to strong coupling asymptotically. This is a direct consequence
of the nontrivial kinetic term for $z$ in the action~(\ref{S4})
and it is a feature that one would have missed had one - inconsistently -
studied the brane-evolution in a static geometry. 

Further, for all solutions the three-brane moves for a finite coordinate
distance only. As a consequence, the brane may or may not collide with one of
the boundaries, depending on initial conditions. The negative-time
branch of these solution constitutes the correct starting point for
moving-brane pre-big-bang cosmology which may terminate in a brane
collision. Whether or not this collision will turn pre-big-bang
contraction of the universe into expansion, thereby generating
a ``graceful exit'', must currently be viewed as an open problem.

The five-dimensional origin of these four-dimensional moving
brane solution is not known at present except for a number of special
cases. There are two four-dimensional solutions with $\e =$ const
which can be directly lifted to five dimensions~\cite{Copeland:2001zp}.
A less trivial example of a five-dimensional moving brane solution
which reduces to one of the above $D=4$ solutions in a specific 
limit only has been found more recently~\cite{Copeland:2002fv}. 

\section{Modelling three-branes with kinks}

Studying more complicated dynamics of three-branes, such as brane-brane
or brane-boundary collision, requires a more microscopic understanding
of M-theory which is currently out of reach. However, these processes
can be studied in the context of a toy model where the three-branes
are replaced by smooth kink solutions of a bulk scalar field theory.
The following action~\cite{Antunes:2002hn} for this toy model,
replacing the M-theory action~(\ref{S5}),
\begin{eqnarray}
 \tilde{S}_5 &=& -\frac{1}{2\k_5^2}\left\{\int_{M_5}\sqrt{-g}\left[
                 \frac{1}{2}R+\frac{1}{4}\partial_\a\F\partial^\a\F
                 +\frac{1}{2}e^{-\F}\partial_\a\c\partial^\a\c
                 +V(\F ,\c )\right]\right.\nonumber \\
             &&\qquad\left. +\int_{M_4^1}\sqrt{-g}\; 2W
                            -\int_{M_4^2}\sqrt{-g}\; 2W\right\}\; .
 \label{S5t}
\end{eqnarray}
has been proposed recently. Here, the potential $V$ is obtained from a
``superpotential'' $W=e^{-\F}w(\c )$ by
\begin{equation}
 V = \frac{1}{3}e^{-2\F}w^2+\frac{1}{2}e^{-\F}U\; ,\qquad
 U = \left(\frac{dw}{d\c}\right)^2\; . \label{U}
\end{equation}
The potential $U$ for the new bulk field $\c$ is required to be
periodic with period $v$, that is $U(\c +v)=U(\c)$ and to have
minima for all $\c =nv$, where $n\in{\bf Z}$, with $U(nv)=0$.
These conditions can be easily translated into conditions on the
function $w(\c )=\int_0^\c d\tilde{\c}\sqrt{U(\tilde{\c})}$ which
specifies the superpotential. An example for $U$ is provided by
the sine-Gordon potential
\begin{equation}
 U = m^2\left[1-\cos\left(\frac{2\p\c}{v}\right)\right]\; ,
\end{equation}
although the specific form of the potential is largely irrelevant
for the subsequent discussion as long as the above general properties
are satisfied.

How does~(\ref{S5t}) relate to the original M-theory action~(\ref{S5})?
Consider first a situation where the field $\c$ is constant throughout
space-time and is located in the $n^{\rm th}$ minimum of $U$, that is,
$\c = nv$. Eq.~(\ref{S5t}) then reduces to the M-theory action~(\ref{S5})
without a three-brane, corresponding to a charge configuration
\begin{equation}
 (\b_1,\b_2,\b_3)=(n,-n,0)\label{c1}\; ,
\end{equation}
if the identification $\s = w(v)$ is being used.

The equations of motion derived from the action~(\ref{S5t}) have
Bogomol'nyi-type first integrals. These first order equations can be
used to obtain a BPS single-kink solution which interpolates between
$\c=nv$ on the first boundary and $\c =(n+1)v$ on the second boundary.
In a thin-kink approximation the $3+1$--dimensional surface defined
by the core of the kink can be described by a Nambu-Goto action and can,
therefore, be identified with the three-brane in Eq.~(\ref{S5}). Hence,
in the background of a single kink, the action~(\ref{S5t}) reduces to
the M-theory action with a single-charged three-brane, corresponding
to charges
\begin{equation}
 (\b_1,\b_2,\b_3)=(n,-(n+1),1)\; .\label{c2}
\end{equation}
Note that different configurations of the scalar field $\c$ in the
defect model correspond to topologically distinct M-theory models.  It
has been shown~\cite{Antunes:2002hn} that a collision of a
kink with a boundary generically leads to a transition between
a vacuum state~(\ref{c2}) with a single-charged three-brane and
a vacuum state~(\ref{c1}) without a three-brane. Topological transitions on the
M-theory side can, therefore, be modelled by simple scalar field
dynamics using the action~(\ref{S5t}). This opens up the possibility
of modelling the evolution of more complicated configurations such as
networks of three-branes in the early universe.


\section*{Acknowledgements} The author is supported by a PPARC Advanced
Fellowship.


\end{document}